\documentclass[twocolumn, 
                          showpacs, 
                          nofootinbib, 
                          nobibnotes, 
                          aps,  
                          amssymb, 
                          amsmath, 
                          floatfix, 
                          reprint, 
                          pra, 
                          notitlepage, 
                          showkeys, 
                          10pt]{revtex4-1}

\usepackage{amssymb}
\usepackage{amsmath}
\usepackage{graphicx}
\usepackage{dcolumn}
\usepackage[colorlinks=true,breaklinks=true,allcolors=blue]{hyperref}
\usepackage{url}
\usepackage{todonotes}
\usepackage{tikz}
\usepackage{dsfont}
\usepackage{epsfig}
\usepackage{color}

\usetikzlibrary{decorations.text}

\usetikzlibrary{shapes.symbols, shapes.arrows, calc, shapes}

\def\equationautorefname~#1\null{Equation~(#1)\null}

\newcommand{\eq}[1]{(\ref{eq:#1})}
\newcommand{\Eq}[1]{Eq.\,\eqref{eq:#1}}

\newcommand{\Fig}[1]{Fig.~\ref{fig:#1}}
\newcommand{\fig}[1]{\ref{fig:#1}}

\newcommand{\App}[1]{App.~\ref{app:#1}}

\newcommand{\1}{\uparrow}

\newcommand{\ket}[1]{|#1\rangle}

\makeatletter
\let\cat@comma@active\@empty
\makeatother

\begin{document}

\title{Quenches near Ising quantum criticality as a challenge for artificial neural networks}

\author{Stefanie Czischek}
\author{Martin G\"arttner}
\author{Thomas Gasenzer}
\affiliation{Kirchhoff-Institut f\"ur Physik, Ruprecht-Karls-Universit\"at Heidelberg, Im Neuenheimer Feld 227, 69120 Heidelberg, Germany}
\date{\today}

\begin{abstract}
The near-critical unitary dynamics of quantum Ising spin chains in transversal and longitudinal magnetic fields is studied using an artificial neural network representation of the wave function.
A focus is set on strong spatial correlations which build up in the system following a quench into the vicinity of the quantum critical point.
We compare correlations observed following reinforcement learning of the network states with analytical solutions in integrable cases and tDMRG simulations, as well as with predictions from a semi-classical discrete Truncated Wigner analysis. 
While the semi-classical approach excells mainly at short times and for small transverse fields, the neural-network representation provides accurate results for a much wider range of parameters. 
Where long-range spin-spin correlations build up in the long-time dynamics we find qualitative agreement with exact results while quantitative deviations are of similar size as for the semi-classically predicted correlations, and slow convergence is observed when increasing the number of hidden neurons.
\end{abstract}

\maketitle

\textit{Introduction.}
Simulating the dynamics of interacting quantum many-body systems out of equilibrium is, in general, a hard problem for classical computers due to the exponential scaling of the Hilbert space dimension with the number of particles. 
Only in a limited number of cases, the quantum dynamics can be solved analytically. 
Such exact solutions exist mostly in one spatial dimension, such as for the transverse-field Ising model (TFIM) \cite{Pfeuty1970a, Calabrese2012a, Calabrese2012b}.
For one-dimensional (1D) systems, also matrix-product-state (MPS) representations of quantum states have proven most useful, including the time-dependent density-matrix renormalization-group (tDMRG) and related methods \cite{White1992, Schollwoeck2011, Vidal2004a, Daley2004a, Sharma2015, Haegeman2016}. 
The tDMRG approach makes use of the fact that, for short-range interactions, an initially unentangled state develops entanglement only gradually and can thus be represented in an efficient way. 
However, for long times or spatial dimensions larger than one ($d>1$), efficient and widely applicable numerical methods for calculating the dynamics are essentially absent.

A key to devising such methods is to know how to efficiently represent the quantum states under consideration.
A novel idea is to make use of the capabilities of machine learning algorithms and artificial neural network (ANN) representations to efficiently exploit the structure of quantum states in a way similar to pattern recognition in image processing  \cite{Biamonte2017, Dunjko2017}.
Such approaches have been applied successfully in various areas of science such as computer vision, natural language processing, as well as in the natural sciences. 
Examples include quantum control \cite{Zahedinejad2016, August2017}, phase classification and recognition in statistical physics \cite{Nieuwenburg2017, Carrasquilla2017}, gravitational wave analysis \cite{Biswas2013} and black hole detection \cite{Pasquato2016} in astronomy, and error correction \cite{Torlai2017, Baireuther2018} in quantum information science.
Recently, it has been proposed to represent quantum states by means of neural networks, providing new ways to compute ground states as well as unitary time evolution  \cite{Carleo2017a, Saito2017, Cai2018}, see also \cite{Deng2017a, Deng2017b, Nomura2017, Kaubruegger2017, Teng2017, Gao2017a, Freitas2018, Carleo2018}, directly related to certain types of tensor network states \cite{Huang2017a, Glasser2018, Clark2018, Chen2018}.

\begin{figure}
\includegraphics[width=\linewidth]{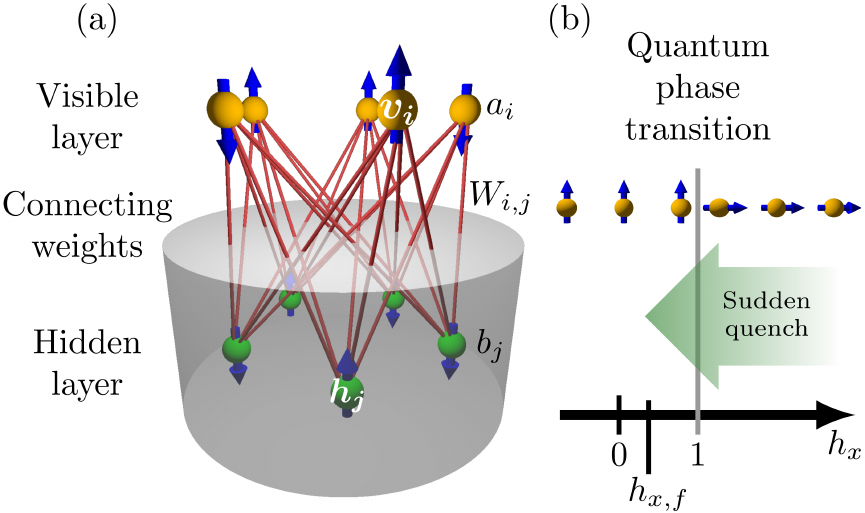}
\caption{(a) Setup of the ANN consisting of one visible and one hidden layer connected via weights $W_{i,j}$ between each pair of variables from different layers and bias $a_i$ (visible) or $b_j$ (hidden) for each variable. (b) Quench protocol in the TFIM for sudden quenches from the deep paramagnetic regime (large $h_{x,i}$) to different distances from the QCP within the paramagnetic and into the ferromagnetic phase, where the ground state configurations are depicted by the spins at the top.}
\label{fig:Fig1}
\end{figure}
%
Here we provide a study of the regimes of validity of the ANN representation of a 1D quantum Ising chain in transverse and longitudinal fields and their relation to the extent of spatial correlations and many-body entanglement, focusing on time evolution after a quench into the vicinity of a quantum critical point.
We furthermore compare capabilities of the ANN method to the results obtained with a discrete truncated Wigner approximation (dTWA) approach recently proposed for quantum spin systems. 
In order to assess the power and weaknesses of these approaches, we make use of analytical solutions for the TFIM, and of numerical simulations by means of exact diagonalization and tDMRG in the case where a longitudinal field breaks the integrability. 

The TFIM  features a quantum phase transition where volume-law growth of entanglement is expected at long times, making it inaccessible to MPS based methods. 
We find that, while the semi-classical method fails to accurately capture the unitary dynamics except at very short times and near zero transverse field, the ANN approach reproduces the exact results for a wide range of parameters.
We show that in cases where strong long-range spin-spin correlations build up, when the system is quenched into the vicinity of a quantum critical point, the two-layer ANN representation requires a strongly increased number of network parameters.
Interestingly, the ANN approach yields qualitatively correct results even where the entanglement entropy increases indefinitely, only limited by the finite system size.
Quantitative deviations, however, are of similar size as in semi-classical predictions and highlight the challenge for the employed ANN approach.

\textit{1D Ising model in longitudinal and transverse fields.}
We consider the dynamics of a periodic chain of $N$ spins governed by the quantum Ising model in a longitudinal field $h_z$ and transverse field $h_x$, with Hamiltonian
\begin{align}
\mathcal{H}=
-\sum_{i=1}^N\sigma^z_i\sigma^z_{\left(i+1\right)\mathrm{mod}N}
-h_x\sum_{i=1}^N\sigma^x_i-h_z\sum_{i=1}^N\sigma^z_i
\,,
\label{eq:HTFIM}
\end{align}
defined in terms of the Pauli matrices $\sigma^\alpha_i$. 
For $h_z=0$ this model reduces to the transverse-field Ising model (TFIM). 
The TFIM is integrable as it can be mapped to free fermions and thus allows for comparisons with exact analytical solutions both, for ground states and unitary time evolution after a parameter quench \cite{Pfeuty1970a,Calabrese2012a,Calabrese2012b,Karl2017a}.
The spin system undergoes a quantum phase transition at $h_{x,c}=\pm 1$, from a ferromagnetic ($0<|h_x|<1$) to a paramagnetic ($|h_x|>1$) phase, as depicted in Fig.~\ref{fig:Fig1}(b). 
For $h_z\neq0$ the model is no longer integrable and does not show a quantum phase transition as the $Z_2$ symmetry is broken explicitly \cite{Ovchinnikov2003}.

\textit{Artificial-Neural-Network approach.}
%
The quantum state of $N$ Ising spins can be expressed in terms of the $2^N$ complex coefficients $c_{\boldsymbol{v}}$, specifying the amplitudes with respect to the basis states $\ket{\boldsymbol{v}}=\ket{v_1,\ldots,v_N}$ ($v_i\in \{-1,+1\}$), i.e. $\ket{\Psi}=\sum_{\boldsymbol{v}} c_{\boldsymbol{v}}\ket{\boldsymbol{v}}$. 
In Ref.~\cite{Carleo2017a}, the representation 
\begin{equation}
 c_{\boldsymbol{v}} = \sum_{\{\mathbf{h}\}} e^{\sum_{i,j} v_i W_{i,j} h_j + \sum_i a_i v_i + \sum_j b_j h_j }
 \label{eq:net}
\end{equation}
of these coefficients in terms of a set of complex parameters $W_{i,j}$, $a_{i}$, $b_{j}$, with $i=1,...,N$, $j=1,...,M$ has been proposed, which is reminiscent of the ANN structure of a restricted Boltzmann machine \cite{Hinton2010}.
This involves an exponential bilinear form of the $N$ visible, i.e., physical spins or neurons $v_{i}$ and of the $M$ hidden or auxiliary classical spin variables (neurons) $h_{j}\in\{-1,+1\}$ [not to be confused with the magnetic fields $h_x$, $h_z$ appearing in Eq.~\eqref{eq:HTFIM}]. The exponential is summed over the hidden spin configurations.
In the following we choose $M=\alpha N$ with integer $\alpha$.
In the two-layer neural network, only connections between the visible and the hidden but not within the layers are allowed, cf.\ \Fig{Fig1}(a). 

To find representations of the form \eq{net} for ground and unitarily evolving states of the model \eq{HTFIM} a variational determination of the complex synaptic weights $W_{i,j}$ and biases $a_i$ and $b_j$ has been proposed in \cite{Carleo2017a}.
This can be interpreted as reinforcement learning of the ANN and is accomplished by means of a stochastic reconfiguration procedure. 
This can be achieved in a numerically efficient way as the sum in the representation \eq{net} can be performed analytically and the configurations $\{\mathbf{v}\}$ of the visible spins can be sampled using Markov-chain Monte Carlo methods. 
For further details of the ANN approach see 
\App{ANN}
and Ref.~\cite{Carleo2017a}.

Note that the number of network parameters $M + N + MN$ scales linearly in the system size $N$ and reduces to $1+\alpha+M$ if translationally invariant solutions are considered \cite{Sohn2012,Carleo2017a}.
Hence, the representation scales polynomially in the size of the system. 
In this respect the method is efficient and similar in spirit to variational Monte Carlo and MPS-based methods \cite{Rubenstein2017, Orus2014, Schollwoeck2011}.
It has been discussed in the context of complexity theory \cite{Gao2017a} and 
exact ANN representations of specific classes of states have been found, including topological cluster and 2D toric-code states \cite{Deng2017b, Gao2017a, Kaubruegger2017}, as well as tensor network and chiral states \cite{Huang2017a, Levine2017a, Glasser2018, Clark2018, Chen2018}.

\textit{Discrete truncated Wigner approximation.}
The dTWA is a semi-classical simulation method for the dynamics of systems defined on a discrete phase space, such as the Ising model defined in \Eq{HTFIM} \cite{Schachenmayer2015, Schachenmayer2015b, Pucci2016, Orioli2017}. 
It allows calculating the dynamics by sampling initial states from a positive definite discrete Wigner function evolved by means of classical equations of motion for the spins. 
By averaging the resulting observables at a given evolution time over a large set of initial-state samples, one finds a semi-classical approximation to the exact unitary evolution (see \App{dTWA} for further details).

\begin{figure*}
\includegraphics[width=\textwidth]{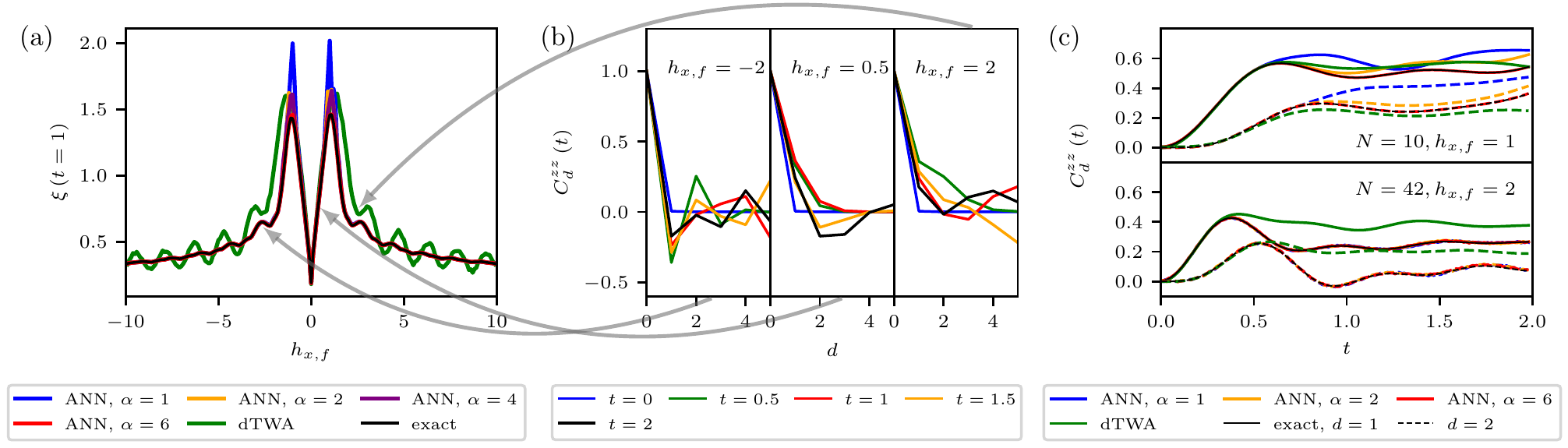}
\caption{(a) Correlation length $\xi\left(t\right)$ at a fixed time $t=1$ as a function of the transverse field $h_{x,f}$ after a sudden quench from $\left(h_{x,i}=100,h_{z,i}=0\right)$ with $h_{z,f}=0$ within the paramagnetic and into the ferromagnetic phase. The results of the ANN approach for different $\alpha$ are compared with the dTWA calculations and the exact solutions. (b) Time dependence of the correlation function $C^{zz}_d\left(t\right)$ after quenches from the deep paramagnetic phase to several $h_{x,f}$. The correlation length in (a) is calculated by fitting an exponential function to the short distance decay of the absolute value of $C^{zz}_d$. (c) Time evolution of the correlation function $C^{zz}_d\left(t\right)$ after a sudden quench to $\left(h_{x,f}=1, h_{z,f}=0\right)$ for a spin chain with $N=10$ sites (upper plot) and to $\left(h_{x,f}=2, h_{z,f}=0\right)$ for a spin chain with $N=42$ sites (lower plot).}
\label{fig:Fig2}
\end{figure*}
%
\textit{Quenches in the TFIM.}
In our analysis we assume the system to be initially in the ground state deep in the paramagnetic phase, $h_{x,i}=100$ and $h_{z,i}=0$, and quench to a range of points $(h_{x,f},h_{z,f})$ in the parameter space of \eq{HTFIM}, cf.~\Fig{Fig1}(b). 
At a given evolution time after the quench we evaluate the spin-spin correlation function
\begin{align}
C^{zz}_d(t)=&\left<\sigma^z_0\sigma^z_{d}\right> \,.
\end{align}
The absolute value of the correlation function typically shows an exponential decay for short distances $d$ between the 
spins. 
We extract a correlation length $\xi$ by fitting $\left|C^{zz}_d\left(t\right)\right|$ with an exponential function $\mathrm{exp}\left(-d/\xi\right)$ at small $d<3$. 

We begin by choosing a zero longitudinal field $h_{z,f}$ and compare the results of the ANN and dTWA methods to exact analytical results for $C^{zz}_d(t)$. 
Restricting ourselves first to small systems  of $N=10$ spins,  Monte Carlo sampling of $\mathbf{v}$ is not necessary in the ANN weight updating step which ensures that effects arising from a finite sample size do not constrain the performance. 
For such small systems, we can furthermore increase the number $M=\alpha N$ of hidden spins up to a size where the number of network parameters exceeds the dimension of Hilbert space and thus an exact ANN representation of the state should be possible. 
In this way we can explore the degree to which the full set of basis states is necessary in representing the system's state.

Our results are presented in \Fig{Fig2}. 
Panel (a) shows the correlation length at a fixed time $t=1$  as a function of $h_{x,f}$. 
\Fig{Fig2}(b) illustrates the $d$-dependence of $C^{zz}_d(t)$ for a selection of $h_{x,f}$ and times $t$, and panel (c) shows the correlation function at two different distances $d$ and final fields $h_{x,f}$ as functions of time after the quench.
Note that the results for $h_{x,f}=2$ in \Fig{Fig2}(c) are for a larger system of $N=42$ spins.

As can be seen in \Fig{Fig2}(a), for quenches into the vicinity of the quantum critical points (QCPs), $h_{x,f}\simeq\pm1$, we observe the buildup of long-range correlations, resulting in strongly increased correlation lengths $\xi$ \cite{Calabrese2012a,Calabrese2012b}. 
While the system possesses a quantum critical point at $h_{x}=\pm1$, i.e., undergoes a quantum phase transition in the ground state, $\xi\left(t\right)$ is not expected to diverge there.
The saturation of the dynamically evolving correlation length is closely related to the fact that the one-dimensional system does not dispose of phase transitions at non-zero temperatures \cite{Karl2017a}.

\begin{figure*}[t]
\includegraphics[width=\textwidth]{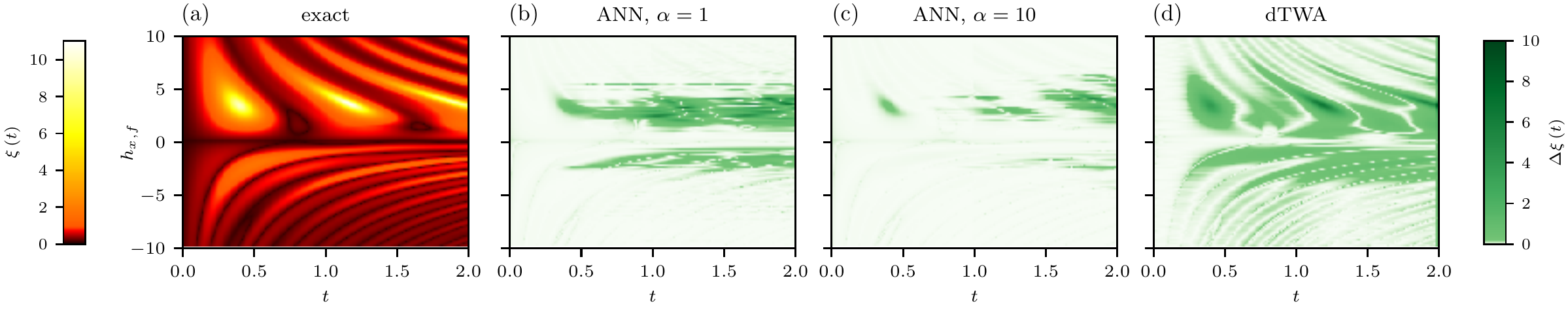}
\caption{(a) Exact correlation length $\xi_e\left(t\right)$ for $N=10$ spins as a function of time after a sudden quench from $\left(h_{x,i}=100,h_{z,i}=0\right)$ to $h_{z,f}=2$ and different values of $h_{x,f}$. (b),(c),(d) Deviations $\Delta\xi\left(t\right)=\left|\xi_s-\xi_e\right|$ of simulation results $\xi_s\left(t\right)$ from the exact result.
The ANN calculation is not limited by Monte Carlo sampling errors. See also \App{LTFIM} for more final values $h_{z,f}$.
}
\label{fig:Fig3}
\end{figure*}

In the vicinity of $h_{x,f}=\pm 1$ the ANN approach, when choosing a small number of hidden spins ($\alpha=1$), yields correlations which deviate clearly from the exact result shown as a black solid line.
Away from these critical values, the obtained results, however, match well with the exact correlations. 
Increasing the number of hidden spins, we find that the accuracy improves, whereas perfect agreement is only obtained for $\alpha>6$. 
Note that then, the number of network parameters is $1+\alpha+M>67$ and thus of the order of the Hilbert space dimension after symmetrization, $d_H=108$.
Around $h_{x,f}=0$, the ANN representation is well controlled as can be shown by means of a perturbative expansion in terms of classical spin networks \cite{Schmitt2017a}.

The time evolution of $C^{zz}_d(t)$ shown in Figure \fig{Fig2}(c), for a quench to $h_{x,f}=1$ and $d=1,2$, corroborates the above findings concerning the dependence on $\alpha$. 
At very short times ($t\lesssim 0.5$), the ANN method gives accurate results already for $\alpha=1$, even for quenches to criticality.
For large spin chains ($N=42$), $\alpha=1$ is still sufficient to capture the exact dynamics in regimes of small correlation lengths ($h_{x,f}=2$ is shown). 
For such large systems sizes, the weight updating procedure requires Monte Carlo sampling of the visible neurons.
This turns out to have no effect other than adding statistical noise to the numerical result, which can be controlled by increasing the size of the Monte Carlo samples. 
By increasing the number of weights up to $\alpha=6$ in the regime of large correlation length, i.e. close to the QCPs, we cannot obtain converged results, which in fact hints to an exponential scaling of the number of required network parameters with $N$.
Hence, in these cases, the method appears to be of no advantage as compared to exact diagonalization (see \App{ANNLarge}).

Turning to the dTWA method we find that it qualitatively reproduces the exact results while, in general, it shows rather large deviations except at short times ($t\lesssim 0.5$) and around $h_{x,f}=0$. 
This is due to the fact that quantum effects are not captured by the semi-classical approximation. 
Note, however, that for quenches very close to the QCPs at $h_{x,f}=\pm1$, the dTWA provides in general a better estimate of the short-range correlation length than efficient ANN representations (see  
\App{TFIM}).

\begin{figure}[t]
\includegraphics[width=\linewidth]{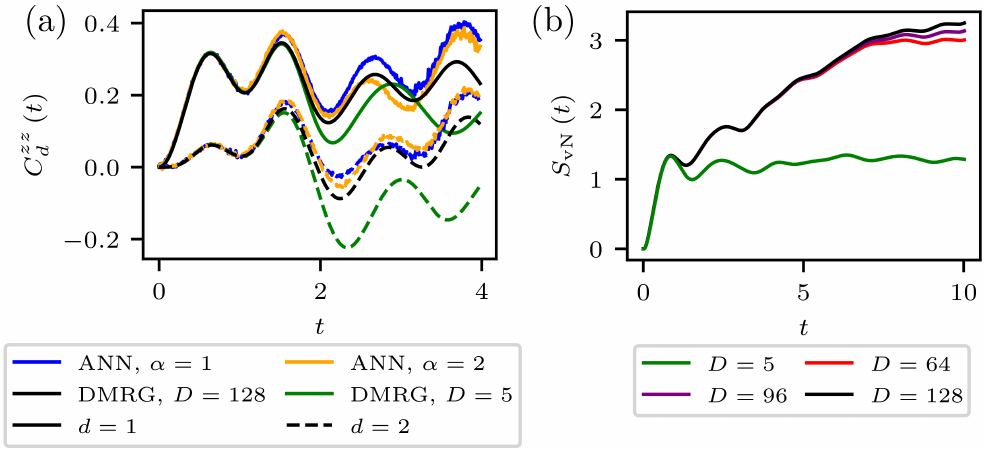}
\caption{(a) Time evolution of the correlation function $C^{zz}_d\left(t\right)$ after a sudden quench from $\left(h_{x,i}=100,h_{z,i}=0\right)$ to $\left(h_{x,f}=0.5,h_{z,f}=1\right)$ in a system with $N=42$ sites. The ANN approach for $\alpha=1$ and $\alpha=2$ is compared to tDMRG calculations with bond dimensions $D=128$ and $D=5$.
(b) von-Neumann entanglement entropy $S_\mathrm{vN}$ as a function of time after the same quench as in (a). The entanglement entropy is calculated using tDMRG with different bond dimensions. The deviations at late times show that in this regime tDMRG breaks down after long times due to the growing entanglement.}
\label{fig:Fig4}
\end{figure}

\textit{TFIM in a longitudinal field.}
Having considered only the integrable TFIM so far, we now turn to the non-integrable case by adding a longitudinal field $h_{z,f}>0$. The initial fields are the same as above, $\left(h_{x,i}=100, h_{z,i}=0\right)$. 
Since no analytical solutions are available in this case, we resort to exact numerical diagonalization for small systems and to the tDMRG.

\Fig{Fig3}(a) shows the exact time evolution of the correlation length after a sudden quench to $h_{z,f}=2$ and different values of $h_{x,f}$ ($N=10$). 
The same dynamics is evaluated using the ANN approach with $\alpha=1$, $\alpha=10$ and the dTWA, where the deviation $\Delta\xi\left(t\right)=\left|\xi_s-\xi_e\right|$ of the correlation length $\xi_s\left(t\right)$ resulting from the different approaches as compared to the exact result $\xi_e\left(t\right)$ is shown in \Fig{Fig3}(b), (c), and (d), respectively. 
The ANN results are close to the exact calculations already for $\alpha=1$ in regimes of small correlation length, namely for $|h_{x,f}|\gg 1$ and $|h_{x,f}|\ll 1$. 
Only for quenches to intermediate $h_{x,f}$, where the correlation length becomes large, deviations can be found for times $t>0.5$, while the first oscillation in the correlation length is captured perfectly even here. 
For $\alpha=10$ only small deviations can be found anywhere. 
In this case, the $111$ weight parameters exceed the Hilbert space dimension $d_{H}=108$, indicating that in certain parameter regimes the number of network parameters needed to achieve full convergence scales exponentially with the system size.
The dTWA [\Fig{Fig3}(d)] shows similiar deviations $\Delta\xi\left(t\right)$ in a wider regime, coinciding with the exact results mainly at short times.

An important question concerns the predictive power of the ANN approach in regimes where MPS-based approaches such as tDMRG are limited to short times due to an extensively growing entanglement entropy.
As a representative case, we show, in \Fig{Fig4}(a), the correlation function after a quench to $\left(h_{x,f}=0.5,h_{z,f}=1\right)$ for a spin chain with $N=42$ sites, where the ANN approach for $\alpha=1$ and $\alpha=2$ is compared to converged tDMRG results for bond dimension $D=128$ and approximate data for $D=5$. 
Clearly, deviations appear for times $t\geq1$, while the qualitative behavior is captured for longer times. 
Increasing $\alpha$ from $1$ ($44$ network parameters) to $2$ ($87$ parameters) does not improve the convergence significantly.
We note that, for $D=128$, the number of variational parameters ($2D^2$) in the MPS vastly exceeds the number of parameters in the ANN approach ($1+\alpha+M$). For bond dimension $D=5$, the tDMRG variational ansatz has $50$ parameters, resulting in a similar quality as the ANN.
\Fig{Fig4}(b) shows the von Neumann entanglement entropy [$S_{\mathrm{vN}}=-\mathrm{Tr}\left(\rho_\mathrm{A}\mathrm{log}\rho_\mathrm{A}\right)$, with half-chain reduced density matrix $\rho_{\mathrm{A}}$] obtained by means of tDMRG with different bond dimensions $D$, as a function of time after the same quench as before. 
Deviations appear at late times, showing that also tDMRG struggles due to the linearly growing entanglement entropy, which in turn requires an exponentially growing bond dimension.

%
\textit{Conclusions.}
Considering sudden quenches in the integrable transverse-field Ising model as well as in the non-integrable Ising model in a transverse and a longitudinal field, we compare the dynamics obtained within the discrete truncated Wigner approximation (dTWA) \cite{Schachenmayer2015,Pucci2016} and a variational approach based on artificial neural networks (ANN) \cite{Carleo2017a} with analytical and exact numerical results.
Both the dTWA and the ANN methods reproduce the short-time dynamics correctly and perform well for small transverse fields. 
The ANN approach agrees with exact calculations in a much wider parameter regime than dTWA. 
Near criticality and at the presently attainable level of accuracy, it shows similar deviations as the semi-classical approach.

\textit{Acknowledgements.} 
The authors thank A.~Baumbach, H.~Cakir, G.~Carleo, J. C. Halimeh, M.~Karl, M.~Kastner, K.~Meier, M.~K.~Oberthaler, J.~M.~Pawlowski, and A.~Pi{\~n}eiro Orioli for discussions and collaboration on the topics described here. The authors thank J.~Schachenmayer for sharing his tDMRG codes. 
This work was supported by the Horizon-2020 framework programme of the European Union (FET-Proactive, AQuS, No. 640800),  by Deutsche Forschungsgemeinschaft (SFB 1225 ISOQUANT), by Heidelberg University, and by the state of Baden-W\"urttemberg through bwHPC.

\bibliographystyle{apsrev4-1}
%



\begin{appendix}
\clearpage
\widetext
\begin{center}
\appendix
\textbf{\large Appendix}
\end{center}
\setcounter{equation}{0}
\setcounter{table}{0}
\setcounter{page}{1}
\makeatletter

\section{The discrete truncated Wigner approximation}
\label{app:dTWA}

The discrete truncated Wigner approximation (dTWA) is based on sampling an initial spin state from a Wigner function on a discrete phase space and classically evolving it in time. By repeating this, one can create lots of trajectories and by averaging the outcoming observables, their time evolution can be approximated semi-classically \cite{Schachenmayer2015, Pucci2016}.

\begin{figure}
\includegraphics[width=0.3\textwidth]{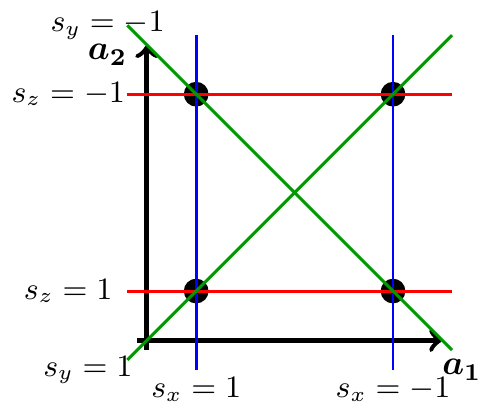}
\caption{Discrete quantum phase space for a spin-$\frac{1}{2}$ system spanned by two variables $a_1$, $a_2$. The colored lines denote the three sets of parallel lines in the $2\times 2$ dimensional finite mathematical field. With each set a spin operator $\sigma^\alpha$, $\alpha\in\left\{x,y,z\right\}$ is associated and each line is identified with one eigenvalue of the corresponding operator.}
\label{SuppFig1}
\end{figure}

A discrete phase space of a quantum spin-$\frac{1}{2}$ system is based on a $2\times 2$ dimensional finite mathematical field, which consists of three sets of parallel lines. As in a continuous phase space, with each set of parallel lines one operator is associated and each line is identified with an eigenvalue of the corresponding operator \cite{Wootters1987}. For a spin-$\frac{1}{2}$ system we associate the spin operators, which are the Pauli operators $\sigma^x$, $\sigma^y$ and $\sigma^z$, with the sets of parallel lines and each line is identified with either the $+1$ or $-1$ eigenvalue, as illustrated in Fig.~\ref{SuppFig1}. For each point $\boldsymbol{\alpha}=\left(a_1,a_2\right)$ in the phase space, a phase point operator $A_{\boldsymbol{\alpha}}$ can be defined, which maps each point in the Hilbert space onto a point in phase space. Here, a convenient choice for the phase point operators is \cite{Schachenmayer2015,Wootters1987}
\begin{align}
A_{\boldsymbol{\alpha}}&=\frac{1}{2}\left[\left(-1\right)^{a_1}\sigma^x+\left(-1\right)^{a_1+a_2}\sigma^y+\left(-1\right)^{a_2}\sigma^z+\mathds{1}\right],
\end{align}
which is consistent with the association of the spin operator eigenvalues in Fig.~\ref{SuppFig1}.

By mapping the density operator $\rho$ onto the phase space, the Wigner function is defined as
\begin{align}
W_{\boldsymbol{\alpha}}&=\frac{1}{2}\mathrm{Tr}\left(\rho A_{\boldsymbol{\alpha}}\right),
\end{align}
which is a quasi-probability distribution over the phase space. This means it gives a probability for each point in phase space and shows properties of a probability distribution, but it might have negative values \cite{Wootters1987,Wigner1932a,Polkovnikov2010a}. By combining two phase spaces with different phase point operators, all eight possible orientations of the discrete spin can be captured and a quasi-probability for each orientation is given \cite{Schachenmayer2015,Pucci2016}.

From the Wigner function on the discrete phase space, initial spin states can be sampled by creating one phase space for each site and sampling the orientation of the corresponding spin. This is only possible if the Wigner function is non-negative, which limits the flexibility of the simulation method. The time evolution of the sampled spin state can then be approximated by treating the quantum spins as classical spins $\boldsymbol{s}_i$ and evolving them in time individually using classical equations of motion \cite{Schachenmayer2015},
\begin{align}
\dot{s}^\alpha_i=&\left\{s^\alpha_i, \mathcal{H}\right\},
\label{DTWA_EOM}
\end{align}
with $\alpha\in\left\{x,y,z\right\}$ and $i\in\left\{1,\dots,N\right\}$ for an N-site spin system. The brackets denote the Poisson bracket, which for two classical spins is defined as
\begin{align}
\left\{s^\alpha_i,s^\beta_j\right\}=&2\delta_{ij}\sum_{\gamma}\epsilon^{\alpha\beta\gamma}s^\gamma_i.
\end{align}
In the dTWA, $R$ classical trajectories are calculated, where the initial state is sampled from the Wigner function for each trajectory individually. The observables resulting from the trajectories are then averaged to approximate quantum dynamics in a semi-classical way, since a classical time evolution is used, but quantum fluctuations come into the system by sampling the initial states.

In the main text, sudden quenches in the transverse field Ising model are considered. The quenches are chosen to start from a large initial transverse field $h_{x,i}=100$, which is sufficiently large to create a fully $x$-polarized ground state, for which the Wigner function is non-negative, making it possible to calculate the dTWA for the considered quenches. To reach more accurate results, we improved the method by trying higher orders in the approximation of the equations of motion \cite{Pucci2016}. For this we observed that taking only the next order into account leads to numerical instabilities, so that even higher orders would be necessary.

\section{The artificial neural network approach}
\label{app:ANN}
The recently introduced artificial neural network (ANN) approach is based on representing a quantum state in terms of the weights in a restricted Boltzmann machine \cite{Carleo2017a}. Using the setup introduced in the main text with $M\times N$ complex connecting weights $W_{i,j}$ between the $N$ visible and $M$ hidden neurons, as well as $N$ visible biases $a_i$ and $M$ hidden biases $b_j$, the wave function $\left|\Psi\right>$ of a quantum spin state can be written as
\begin{align}
\left|\Psi\right>=&\sum_{\boldsymbol{v}}c_{\boldsymbol{v}}\left(\mathcal{W}\right)\left|\boldsymbol{v}\right>,\\
\mathrm{with\ }c_{\boldsymbol{v}}\left(\mathcal{W}\right)=&\sum_{h_j\in\left\{\pm1\right\}}\mathrm{exp}\left(\sum_{i=1}^N\sum_{j=1}^Mv_iW_{i,j}h_j+\sum_{i=1}^Na_iv_i+\sum_{j=1}^Mb_jh_j\right)\\
=&\mathrm{exp}\left(\sum_{i=1}^Na_iv_i\right)\prod_{j=1}^M2\mathrm{cosh}\left(b_j+\sum_{i=1}^Nv_iW_{i,j}\right),
\end{align}
with the vector $\mathcal{W}=\left(a_1,\dots,a_N,b_1,\dots,b_M,W_{1,1},\dots,W_{N,M}\right)$ of all weight variables.

In a general restricted Boltzmann machine with real weights, the expression for $c_{\boldsymbol{v}}\left(\mathcal{W}\right)$ would correspond to the probability assigned to a configuration of visible variables \cite{Hinton2010}, but since we have complex weights here, also $c_{\boldsymbol{v}}\left(\mathcal{W}\right)$ is complex and does not describe a probability. Instead its square corresponds to the probability of the configuration $\boldsymbol{v}$, as $c_{\boldsymbol{v}}\left(\mathcal{W}\right)$ is the prefactor of the product state $\left|\boldsymbol{v}\right>$ in the wave function.

With this representation, the ground state wave function of a quantum spin-$\frac{1}{2}$ system can be found using an iterative scheme based on a Stochastic Reconfiguration method \cite{Sorella2007,Sorella2001}. The weight parameters $\mathcal{W}$ at iteration step $p+1$ can be calculated using the vector of forces $F$ and the covariance matrix $S$ \cite{Carleo2017a},
\begin{align}
\mathcal{W}\left(p+1\right)=&\mathcal{W}\left(p\right)-\gamma S^{-1}\left(p\right)F\left(p\right),\\
\mathrm{with\ }S_{kk'}=&\left<\mathcal{O}_k^*\mathcal{O}_{k'}\right>-\left<\mathcal{O}^*_k\right>\left<\mathcal{O}_{k'}\right>,\\
F_k\left(p\right)=&\left<E_{\mathrm{loc}}\mathcal{O}^*_k\right>-\left<E_{\mathrm{loc}}\right>\left<\mathcal{O}^*_k\right>,
\end{align}
with an iteration step size $\gamma$ and the star denoting complex conjugation. To calculate the inverse of $S$ in a stable way, a regularization method is used \cite{Carleo2017a}. The local energy $E_{\mathrm{loc}}$ and the variational derivative $\mathcal{O}_k$ are defined as
\begin{align}
E_{\mathrm{loc}}\left(\boldsymbol{v}\right)=&\frac{\left<\boldsymbol{v}\right|\mathcal{H}\left|\Psi\right>}{c_{\boldsymbol{v}}},\\
\mathcal{O}_k\left(\boldsymbol{v}\right)=&\frac{1}{c_{\boldsymbol{v}}}\partial_{\mathcal{W}_k}c_{\boldsymbol{v}}.
\end{align}
Here, the Hamiltonian $\mathcal{H}$ of the quantum spin system under consideration enters into the algorithm. The expectation values are generally calculated as
\begin{align}
\left<\mathcal{O}\right>=&\left<\Psi\right|\mathcal{O}\left|\Psi\right>\\
=&\sum_{\boldsymbol{v},\tilde{\boldsymbol{v}}}\left<\boldsymbol{v}\right|\mathcal{O}\left|\tilde{\boldsymbol{v}}\right>c_{\tilde{\boldsymbol{v}}}c_{\boldsymbol{v}},
\end{align}
where the summations run over all states in the Hilbert space, which are $2^N$ states for an $N$-site spin system.

Since such a large number of states can not be taken into account for large spin systems, it is more convenient to use only a subset of states, which is generated by a Markov chain set up by sampling $\left|c_{\boldsymbol{v}}\right|^2$ using a Metropolis-Hastings algorithm \cite{Carleo2017a}. The Markov chain is created starting from a random initial state $\boldsymbol{v}^k$ by flipping a random spin to get a new configuration $\tilde{\boldsymbol{v}}$. The new configuration is accepted with probability $A\left(\boldsymbol{v}^k\rightarrow\tilde{\boldsymbol{v}}\right)=\mathrm{min}\left(1,\left|c_{\tilde{\boldsymbol{v}}}/c_{\boldsymbol{v}^k}\right|^2\right)$. If accepted, the configuration is updated, $\boldsymbol{v}^{k+1}=\tilde{\boldsymbol{v}}$, if rejected, the state stays the same, $\boldsymbol{v}^{k+1}=\boldsymbol{v}^k$. This way, the Markov chain is set up in an iterative scheme and in the end creates a subset of states with large coefficients $\left|c_{\boldsymbol{v}}\right|^2$. Since these states have high contributions to the expectation values, this gives a reasonable approximation, since states with smaller contributions can be neglected. This technique is used routinely in variational Monte Carlo and is known to be stable and efficient.

For small system sizes, as considered in the main text, this sampling procedure is not necessary, since the expectation values can be calculated exactly due to the small dimension of the Hilbert space.

The iteration scheme to find the ground state can be interpreted as an effective imaginary time evolution and in an analogous way an iteration scheme to approximate the real time evolution of the quantum spin system can be derived using a time-dependent variational Monte Carlo approach \cite{Carleo2017a,Carleo2014,Carleo2012}. The equations of motion for the weight parameters $\mathcal{W}$ again depend on the vector of forces $F$ and the covariance matrix $S$, which are defined in the same way as above and give
\begin{align}
\dot{\mathcal{W}}\left(t\right)=&-iS^{-1}\left(t\right)F\left(t\right).
\end{align}
These can be easily integrated numerically. Here, $S^{-1}$ denotes the Moore-Penrose pseudo-inverse of $S$, which is not necessarily invertible since it is not guaranteed to have full rank. To get stable results even for large spin systems and in the vicinity of a quantum critical point, we found that the pseudo-inverse needs to be combined with the regularization method used in the ground state calculation.

Given this, the simulations of the dynamics after sudden quenches as considered in the main text can be calculated by starting with random weights, converging to the initial ground state and calculating the time evolution starting from the ground state weights using the parameters after the quench. 

By introducing symmetries into the ANN setup, the number of weight parameters can be reduced. In the case of spin chains with periodic boundary conditions, as discussed in the main text, we find translation invariance. To include this into the ANN approach, we shift all visible variables by a factor $d$ around the ring and force the new configuration to have the same weight parameters as the old configuration. Doing so reduces the number of weight parameters from $M+N+MN$ to $M+\frac{M}{N}+1$ \cite{Carleo2017a,Sohn2012}.

\section{Transverse Field Ising Model}
\label{app:TFIM}
The one-dimensional transverse field Ising model (TFIM) with $N$ spin-$\frac{1}{2}$ sites is described by the Hamiltonian
\begin{align}
H=&-J\sum_{i=1}^{N}\sigma^z_i\sigma^z_{\left(i+1\right)\mathrm{mod}N}-h_x\sum_{i=1}^N\sigma^x_i,
\end{align}
with the Pauli matrices $\sigma^\alpha_i$, and we choose $J=1$ without loss of generality. This model is integrable, since it can be mapped onto non-interacting fermions, from which spectrum and energy eigenstates can be calculated analytically \cite{Pfeuty1970a,Calabrese2012a,Calabrese2012b}. The model shows a quantum phase transition at $h_c=\pm1$ between a paramagnetic ($\left|h_x\right|>1$) and a ferromagnetic ($\left|h_x\right|<1$) phase. This quantum phase transition is quantified by a gapless dispersion relation \cite{Calabrese2012a,Calabrese2012b,Karl2017a}. It has been shown that for sudden quenches from a large transverse field into the vicinity of the quantum critical point, as we are considering them in the main text, the correlation length shows the behavior of a generalized Gibbs ensemble (GGE), so it increases for smaller distances from the quantum critical point \cite{Calabrese2012a,Calabrese2012b,Karl2017a}.

\begin{figure}
\includegraphics{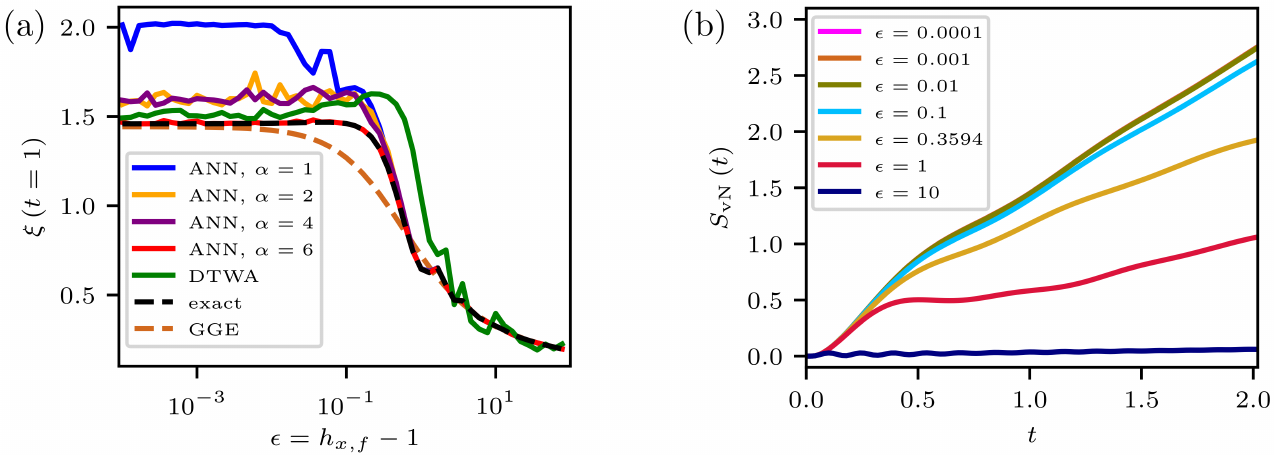}
\caption{Correlation length at a fixed time (a), time evolution of the von-Neumann entanglement entropy (b) after a sudden quench from $\left(h_{x,i}=100,h_{z,i}=0\right)$ to different distances $\epsilon$ from the quantum critical point in the TFIM with $h_{z,f}=0$. In the region of large correlation lengths, large $\alpha$ are needed in the ANN approach to capture the exact solution. In the same region also volume-law entanglement is found.}
\label{Fig_S2}
\end{figure}
Fig.~\ref{Fig_S2}(a) shows this behavior of the correlation length at a fixed time $t=1$ after quenches within the paramagnetic phase for a spin chain with $N=10$ sites, where the distance $\epsilon=\left(h_{x,f}-h_c\right)/h_c$ from the quantum critical point is plotted on a logarithmic scale. The different simulation methods are compared with the exact solution of the model. The correlation length is here extracted from the equal-time correlation function $C^{zz}_d\left(t\right)$ by fitting an exponential function to the short distance decay of $C^{zz}_d\propto\exp\left(-d\xi^{-1}\right)$ for small $d$. The solid line shows the GGE behavior, which is given by \cite{Calabrese2012a,Calabrese2012b}
\begin{align}
\xi_{GGE}\left(\epsilon\right)=&\frac{1}{\mathrm{ln}\left(2\left(\epsilon h_c+h_c\right)\right)}.
\end{align}
The exact solution describes the GGE behavior well except for the region around $\epsilon=10^{-1}$. Here, the exact solution is not yet saturated and a longer evolution time is needed until it is converged onto the GGE curve. Since we are only considering systems with $N=10$ sites here, the correlation length does not saturate before finite size effects appear, so the GGE curve is never reached completely. As already discussed in the main text, the ANN approach needs larger $\alpha$ to capture the exact solution in the regime of large correlation lengths, while it works very well even for $\alpha=1$ at small correlation lengths. This behavior can also be observed in Fig.~\ref{Fig_S2}(a). It can also be seen that the dTWA is closer to the exact solution in the vicinity of the QCP, but gets worse when the transverse field increases, as discussed in the main text.

Fig.~\ref{Fig_S2}(b) shows the time evolution of the von-Neumann entanglement entropy, which we calculate using tDMRG. There one can see that for $h_{x,f}<2$ the entropy grows linearly with time, while it stays constant for $h_{x,f}\geq2$. To better see this linear growth, the tDMRG calculations are done for a chain with $N=40$ sites, so that finite size effects do not appear on this time scale. Here one can directly see that in the regime where large $\alpha$ are necessary in the ANN approach, not only the correlation length is large, but also volume law entanglement is found. This volume law entanglement also limits tDMRG calculations in this regime for large spin chains and longer times, so that the ANN approach is limited in the same regime as the tDMRG calculations in the TFIM.

If we quench the transverse field to $h_{x,f}<0$, we again find a gapless dispersion relation at $h_{c,2}=-1$, where the gap now closes at the edge of the Brioullin zone, while it closes in the middle for $h_c=1$. Hence, we find another quantum phase transition between a paramagnetic ($h_{x,f}<-1$) and a ferromagnetic ($h_{x,f}>-1$) phase. In this regime, the nearest-neighbor correlation function gets negative. The correlation length can then be extracted from the absolute value of the correlation function in the same way as before, which results in the symmetric curve around $h_{x,f}=0$ plotted in the main text in Fig.~\ref{fig:Fig2}(a).

\section{Ising Model in Transverse and Longitudinal Field}
\label{app:LTFIM}
If a longitudinal field $h_z$ is added to the TFIM, the Hamiltonian becomes
\begin{align}
H=&-J\sum_{i=1}^N\sigma^z_i\sigma^z_{\left(i+1\right)\mathrm{mod}N}-h_x\sum_{i=1}^N\sigma^x_i-h_z\sum_{i=1}^N\sigma^z_i,
\end{align}
and the model is not integrable any more. Again we can choose $J=1$ without loss of generality and we again find a paramagnetic phase for large $h_x$ and small $h_z$ and a ferromagnetic phase for small $h_x$. The difference to the TFIM is now that there is no quantum phase transition between the phases, since the ground state in the ferromagnetic regime is not degenerate anymore due to the longitudinal field and hence there is no spontaneous symmetry breaking between the phases \cite{Ovchinnikov2003}.

\begin{figure}
\includegraphics[width=\textwidth]{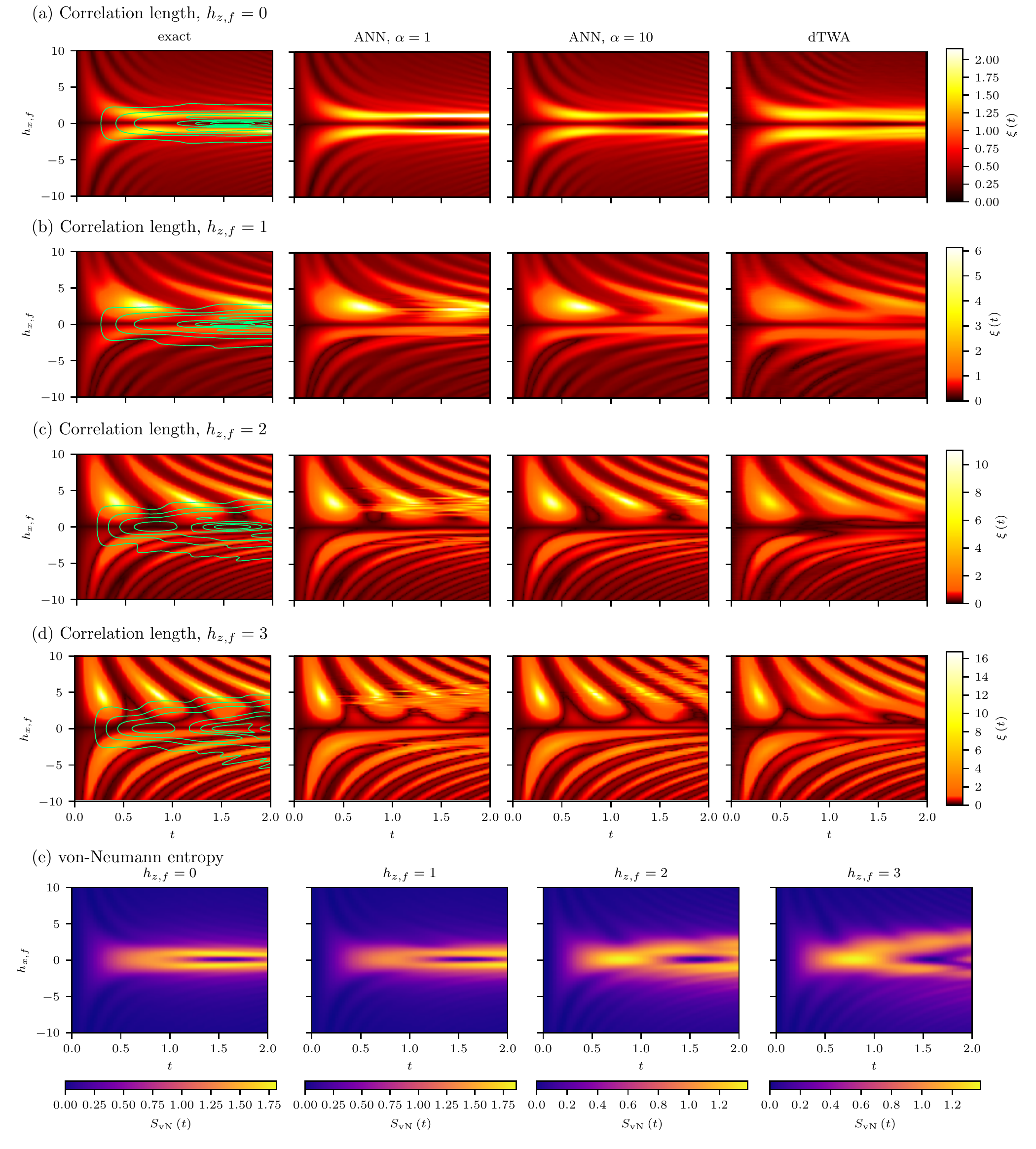}
\caption{Time evolution of correlation length and von-Neumann entropy after sudden quenches from $\left(h_{x,i}=100, h_{z,i}=0\right)$ to different values of $h_{x,f}$ with $h_{z,f}=0$ (a), $h_{z,f}=1$ (b), $h_{z,f}=2$ (c) and $h_{z,f}=3$ (d). The ANN approach with $\alpha=1$ and $\alpha=10$ as well as the dTWA are compared with exact results. (e) shows the time evolution of the von-Neumann entanglement entropy extracted from the tDMRG calculations for $h_{z,f}=1$, $h_{z,f}=2$ and $h_{z,f}=3$. The entanglement entropy is also plotted as contours in the color plot of the exact correlation length in (a), (b), (c) and (d). There one can directly see that large entanglement entropy and volume-law entanglement can be found in the same regimes as large correlatin lengths.}
\label{Fig_S3}
\end{figure}
Considering now quenches from $\left(h_{x,i}=100, h_{z,i}=0\right)$ to different $h_{x,f}$ and $h_{z,f}$, the spin dynamics show Rabi oscillations due to the interaction of $h_x$ and $h_z$. These oscillations can also be found in the correlation length extracted from the equal-time correlation function $C^{zz}_d\left(t\right)$ in the same way as discussed earlier. The time evolution of the correlation length is plotted in Fig.~\ref{Fig_S3} for quenches to different $h_{x,f}$ with $h_{z,f}=0$ (a), $h_{z,f}=1$ (b), $h_{z,f}=2$ (c) and $h_{z,f}=3$ (d). We compare the simulation results with exact diagonalization calculations, since the model is not analytically solvable anymore. We can observe in all plots that the ANN approach shows fluctuations for $\alpha=1$ at times $t>0.5$ in the regime where the correlation lengths get large at shorter times. For $\alpha=10$, the method can capture the exact solution quite well, only small deviations appear at very late times. It is interesting to see that even for $\alpha=1$ the first oscillation is always captured perfectly, it only breaks down at later oscillations. For $h_{z,f}=0$ deviations in the $\alpha=1$ calculations only appear in the regime of large correlation length, where the simulations reach too large values. This is directly at the quantum phase transition at $h_{x,f}=1$, where also volume-law entanglement can be found.

In Fig.~\ref{Fig_S3}(e) the time evolution of the von-Neumann entanglement entropy is shown for $h_{z,f}=1$, $h_{z,f}=2$ and $h_{z,f}=3$, which is calculated using tDMRG. Here we can see that the entanglement entropy gets large approximately in the same regime as the correlation length, which is also where deviations in the ANN approach for small $\alpha$ can be found.

The dTWA also captures the oscillations in the correlation length, but at later times deviations can be found for all $h_{x,f}$. The maximum values of the correlation length are never reached in these calculations and deviations are found even at shorter times than in the ANN approach.

\section{Artificial Neural Network Approach for Large Spin Systems}
\label{app:ANNLarge}
\begin{figure}
\includegraphics[width=\textwidth]{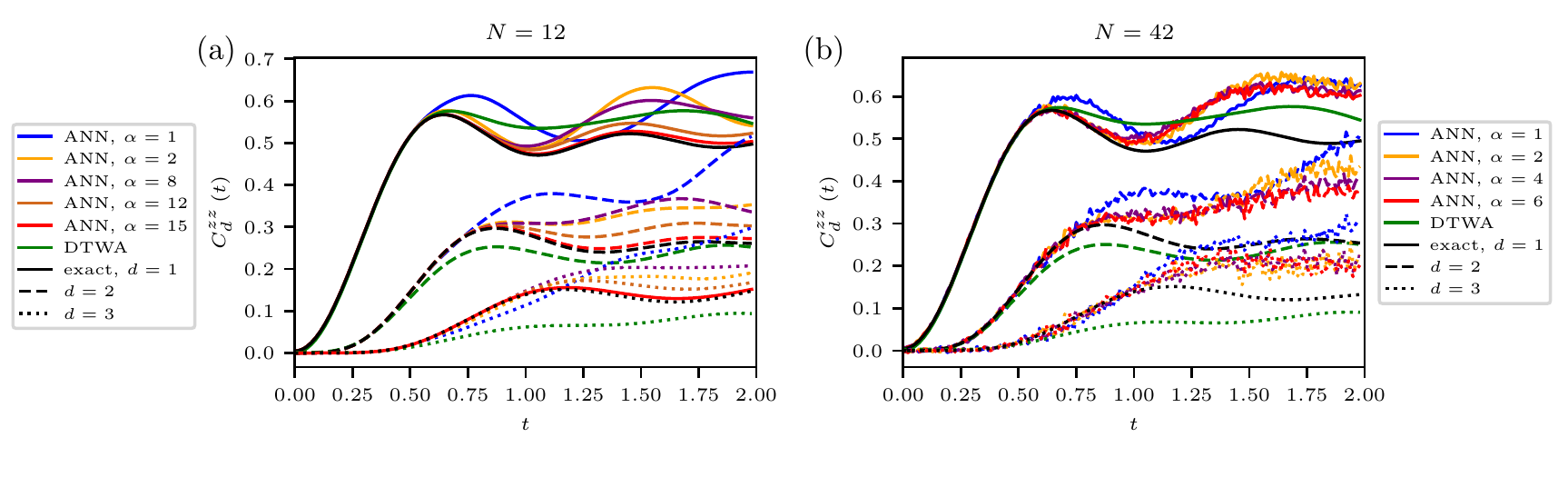}
\caption{Time evolution of the correlation function $C^{zz}_d\left(t\right)$ after a sudden quench from $\left(h_{x,i}=100,h_{z,i}=0\right)$ to $\left(h_{x,f}=1,h_{z,f}=0\right)$ in a spin chain with $N=12$ (a) and $N=42$ (b) sites. The correlation function is shown for distances $d=1$ (solid lines), $d=2$ (dashed lines) and $d=3$ (dotted lines) between the considered spins. The number of weight parameters in the ANN approach is increased to see that $\alpha=15$ is necessary to capture the exact dynamics for $N=12$, while full convergence can not be reached for $N=42$ within suitable computation time.}
\label{Fig_S4}
\end{figure}
In the main text we mostly considered small spin chains with $N=10$ sites. For these systems we found that in the ANN approach the number of weight parameters needs to be as large as the dimension of the Hilbert space to represent the exact dynamics in regimes of large correlations. To see how the necessary number of weight parameters depends on the system size, we consider the quench from $\left(h_{x,i}=100,h_{z,i}=0\right)$ to $\left(h_{x,f}=1,h_{z,f}=0\right)$ and compare the ANN and dTWA simulations to exact solutions for spin chains with $N=12$ and $N=42$ sites. For $N=12$, the number of weight parameters can still be increased until the Hilbert space dimension is reached, while this is not possible anymore for $N=42$. Also the Monte Carlo sampling needs to be used for the $N=42$ calculations, since a summation over all configurations is not possible.

We have already shown in the main text that even for $N=42$, $\alpha=1$ is sufficient for quenches into regimes of small correlations, but larger $\alpha$ is needed in regimes of larger correlations even if a longitudinal field is added. For the quench we are considering here, we found in the main text that $\alpha=6$ is necessary for $N=10$ sites. Fig.~\ref{Fig_S4} shows the time evolution of the correlation function for $N=12$ (a) and $N=42$ (b). There we find that for $N=12$, $\alpha=15$ is necessary to capture the exact dynamics, which corresponds to $1+\alpha+M=196$ weight parameters. This is again of the order of the Hilbert space dimension after symmetrization ($d_H=352$), as we also found it for $N=10$ sites. This suggests that the necessary number of weight parameters scales exponentially with the system size.

In Fig.~\ref{Fig_S4}(b), we increase $\alpha$ as far as possible within suitable computation time for $N=42$ sites. One can see the convergence to the exact dynamics with increasing $\alpha$, but for $\alpha=6$ the result is still far away from the exact solution. The small fluctuations in the ANN calculations are caused by the finite Monte Carlo sampling. A much larger $\alpha$ would be necessary here. This directly shows the limitations of the ANN approach for large spin systems in regimes of large correlations due to the exponential scaling of $\alpha$ with system size, while we have shown in the main text that the method works perfectly fine for large spin systems and small $\alpha$ in regimes of small correlations.

\end{appendix}

\end{document}